# Transmission of travelling-wave with a simple waveguide for rodents MRI at 9.4 T

F. Vázquez[1], O. Marrufo[2], R. Martin[1], S. Solis[1], A. O. Rodriguez[3]

[1]Department of Physics, Faculty of Sciences, Universidad Nacional Autonoma de Mexico, Mexico DF 04510, Mexico.
[2]Department of Neuroimaging, National Institute of Neurology and Neurosurgery MVS, Mexico DF 14269, Mexico.
[3]Department of Electrical Engineering, Universidad Autonoma Metropolitana Iztapalapa, Mexico DF 09340, Mexico.

**Introduction**

Standard coils can cause inhomogeneities due to the standing wave patterns at ultra high field MRI. This B1 inhomogeneity can be overcome using the travelling-wave MRI (twMRI) approach. The majority of twMRI research has been done with clinical MRI scanners at 7T or above [1]. More recently, this concept has been also used with animal MRI systems [2], using circular-cross section waveguides and dielectric materials for both type of scanners. We have demonstrated that twMRI can be used at 3 T with clinical systems and a parallel-plate waveguide [3]. Here, we investigated the use of a parallel-plate waveguide and a RF circular coil to generate rat images with an animal MRI at 9.4 T.

**Material and Methods**

We developed a waveguide with the simplest layout, using two parallel plates made out of copper strips. The strips were mounted on an acrylic 2 m long tube with a 7 cm diameter. Figure 1.a). shows the waveguide design with one end blocked. A saline solution- filled sphere (4 cm diameter) was used for phantom imaging, and a Sprague Dawley rat weighing 240 g was used for ex vivo imaging experiments. A circular coil (5.8 cm diameter) was used for transmission and reception of the MR signal. To improve transmission of the signal, a cylindrical bazooka was added and tuned to 400 MHz (resonant frequency for protons at 9.4T). Both of them were positioned at the magnet isocentre as shown in Figure 1. The following acquisition parameters were used; a) phantom imaging: TR/TE=20/4 ms, FA=180 FOV=100x100 mm, matrix size=128x128, thickness= 4 mm, NEX=4, and b) ex vivo imaging: TR/TE=20/4ms, FA=180 FOV=60x60 mm (axial) & FOV=70x70 mm (saggital), matrix size=128x128, thickness= 4mm, NEX=32. All experiments were run in an animal Varian 9.4 T/21 cm MRI system (Agilent, Inc, Palo Alto, CA).

**Results and Discussion**

Phantom and ex vivo rat images were acquired with the parallel-plate waveguide and a circular coil as shown in Figure 2. Phantom images showed good quality with minimal distortions and an acceptable SNR of ~ 6. Rat images showed reasonable image quality with clear delineation of anatomical structures, including the anterior commissure, optic nerve, and olfactory lobe. This is in very good concordance with results obtained at 3T with a clinical whole-body system [3]. These results demonstrated that the parallel-plate waveguide is able transmit the MR signal without any restriction. This also corroborates the theory that cut-off frequency is zero for a parallel-plate waveguide without restrictions of the bore size, allowing us to acquire images at ultra high field MR images for animal studies.

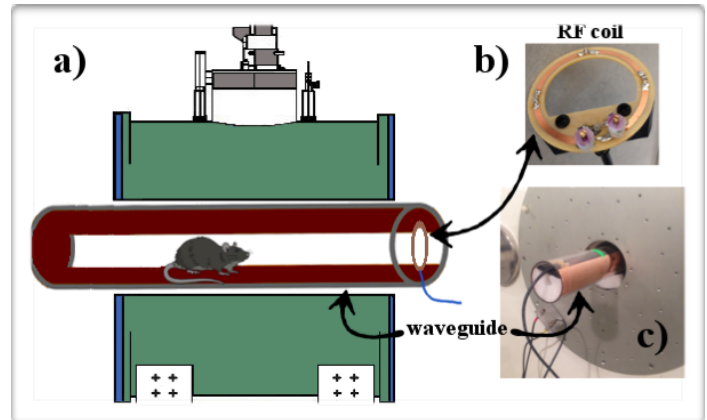

Figure 1. Experimental setup: a) parallel-waveguide inside the magnet bore, b) photo of the circular coil used for transmission and reception of the MR signal, c) photo of RF coil inside the waveguide.

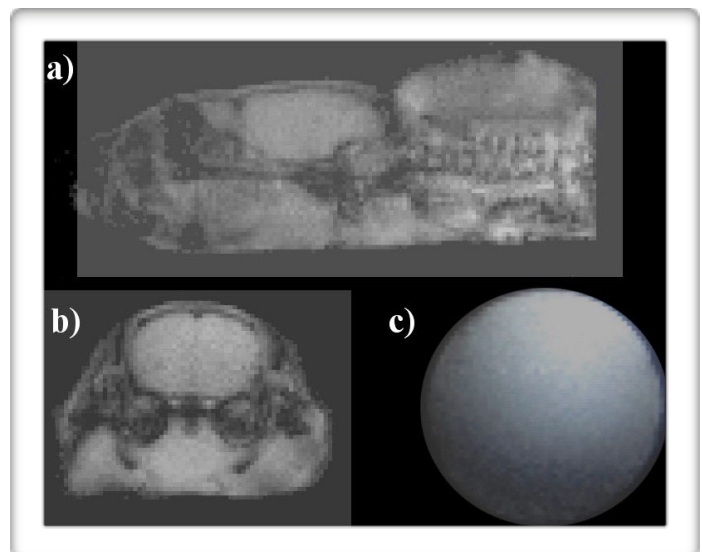

Figure 2. Sprague Dawley rat images: a) saggital whole-body image and b) axial images of the rat's head, c) Mineral spherical phantom image.

**Acknowledgments.**

We thank CONACYT-Mexico for research grant 112092 and Vanderbilt University Institute of Imaging Science (UIIS) for providing technical resources. email: arog@xanum.uam.mx.